\documentclass[conference]{IEEEtran}
\IEEEoverridecommandlockouts
\usepackage{cite}
\usepackage{amsmath,amssymb,amsfonts}
\usepackage{algorithmic}
\usepackage{graphicx}
\usepackage{textcomp}
\usepackage{xcolor}
\usepackage{booktabs}
\usepackage{multirow}
\usepackage{cleveref}
\usepackage{algorithm}
\usepackage{algorithmic}
\usepackage{graphicx}
\usepackage{subcaption}
\usepackage{float}

\def\BibTeX{{\rm B\kern-.05em{\sc i\kern-.025em b}\kern-.08em
    T\kern-.1667em\lower.7ex\hbox{E}\kern-.125emX}}
\begin{document}

\title{Towards Propagation-aware Representation Learning for Supervised Social Media Graph Analytics}

\author{\IEEEauthorblockN{Wei Jiang$^{1}$, Tong Chen$^{1}$, Wei Yuan$^{1}$, Xiangyu Zhao$^{2}$, Quoc Viet Hung Nguyen$^{3}$, Hongzhi Yin$^{1*}$}
\IEEEauthorblockA{\textit{$^{1}$The University of Queensland, Australia} \\
\textit{$^{2}$City University of Hong Kong, Hong Kong}\\
\textit{$^{3}$Griffith University, Australia}\\
\{wei.jiang, tong.chen, w.yuan, h.yin1\}@uq.edu.au, xy.zhao@cityu.edu.hk, henry.nguyen@griffith.edu.au}
\thanks{$^*$Hongzhi Yin is the corresponding author.}
}

\maketitle

\begin{abstract}
Social media platforms generate vast, complex graph-structured data, facilitating diverse tasks such as rumor detection, bot identification, and influence modeling. Real-world applications like public opinion monitoring and stock trading -- which have a strong attachment to social media -- demand models that are performant across diverse tasks and datasets. However, most existing solutions are purely data-driven, exhibiting vulnerability to the inherent noise within social media data. Moreover, the reliance on task-specific model design challenges efficient reuse of the same model architecture on different tasks, incurring repetitive engineering efforts. To address these challenges in social media graph analytics, we propose a general representation learning framework that integrates a dual-encoder structure with a kinetic-guided propagation module. In addition to jointly modeling structural and contextual information with two encoders, our framework innovatively captures the information propagation dynamics within social media graphs by integrating principled kinetic knowledge. By deriving a propagation-aware encoder and corresponding optimization objective from a Markov chain-based transmission model, the representation learning pipeline receives a boost in its robustness to noisy data and versatility in diverse tasks. Extensive experiments verify that our approach achieves state-of-the-art performance with a unified architecture on a variety of social media graph mining tasks spanning graph classification, node classification, and link prediction. Besides, our solution exhibits strong zero-shot and few-shot transferability across datasets, demonstrating practicality when handling data-scarce tasks. The code is available at https://github.com/WeiJiang01/RPRL.
\end{abstract}


\section{Introduction}
With the rapid expansion of the Internet, social media platforms such as X, Facebook and Weibo have become the central hub for information dissemination and public discourse~\cite{kwak2010twitter,chen2011comparision}.
To maximize the value of numerous interactions on these platforms, there is an ongoing trend to model social media data as graphs for a wide range of analyses \cite{jiang2024challenging,yin2019social,hung2017computing,sun2023self,chen2024hate}. Among them, supervised predictive tasks have enjoyed prolonged popularity \cite{liu2024bourne,gao2024accelerating}. Based on the granularity of both graphs and labels, supervised social media graph analytics can be categorized into three core tasks:
(1) graph classification, where one example is rumor detection~\cite{bondielli2019survey} that predicts the veracity of the source information given its propagation graph;
(2)~node classification, such as detecting anomalous users or social bots~\cite{cresci2020decade} within a large user network; and (3) link prediction, where a typical instance is information cascade prediction~\cite{li2017survey}.

While significant progress has been made in each type of these tasks, they are often studied independently, with task-specific models designed around the structural characteristics of their input graphs. For instance, graph classification usually involves small-scale graphs representing information dissemination in a short time period, where models focus on capturing global structural patterns such as centrality and depth~\cite{cui2024propagation,jiang2025epidemiology,sun2021heterogeneous}.
In node classification, the input is typically a single, large-scale social network, and models learn to infer node labels using local neighborhood structures and node attributes~\cite{feng2021botrgcn,liu2023botmoe,yang2024sebot,liu2023imbalanced}.
Link prediction, by contrast, often requires reasoning over dynamic or higher-order structures such as temporal graphs or hypergraphs, demanding more sophisticated models like transformers or hypergraph neural networks~\cite{sun2022ms,qiao2023rotdiff,jiao2024enhancing}.

However, these specialized approaches face several critical limitations.
First, it is notoriously challenging to train effective graph representation learning models with social media data, which is commonly agreed to be rich-yet-noisy \cite{zimmerman2022social,buglass2016friends,jiang2025epidemiology} (e.g., incompleteness, temporal shifts, spurious edges, etc.). This issue is further amplified considering that most existing solutions bear a purely data-driven design \cite{jiang2025epidemiology}.
Second, in real-world applications where social media plays a central role, the combined power of multiple graph tasks \cite{levinson2017government,kavanaugh2011social} is in high demand. For example, political consulting agencies need the insight from diverse social media mining tasks to inform campaign strategy during an election \cite{chauhan2021emergence}, while stock trading firms also clearly benefit from various aforementioned tasks to estimate the public's confidence in certain industries \cite{bukovina2016social}. 
When operated at scale, designing separate models for each task not only incurs substantial engineering efforts, but also risks a high opportunity cost from the repetitive trial-and-error process in the search for optimal model designs.
As such, both factors highlight a critical utility gap between the outgoing, task-specific solutions and practical deployment, and motivate the pursuit of a generic model architecture capable of handling all three task types when trained with corresponding datasets.

Nevertheless, achieving such structural unification remains highly non-trivial. First, the substantial heterogeneity across tasks poses a major challenge, necessitating a generalizable encoder that can effectively model diverse graph structures and feature spaces. Second, learning the dynamics of information propagation is essential for most social media graph mining tasks, yet current methods lack a principled mechanism to describe and quantify these patterns in a consistent and transferable manner across datasets and tasks. In contrast, other domains have demonstrated the potential of integrating physics-informed or first-principle knowledge to guide learning. For instance, traffic forecasting and air quality prediction have successfully incorporated energy-based formulations to model spatiotemporal dynamics through physical laws of energy transfer. These successes further motivate us to explore the use of first principle knowledge to guide representation learning of information propagation in social media networks~\cite{jiang2024physics,hettige2024airphynet}, which can potentially serve as a unifying principle for social media graph learning.

To this end, we design a general architecture that integrates data-driven modeling with principled propagation mechanisms. Concretely, we propose a parallel encoding module that jointly captures graph topology and global context. It consists of two complementary encoders: a structure-aware graph encoder that models local dependencies based on actual connectivity, and a structure-agnostic context encoder that captures global interactions across all nodes regardless of edge information. This design enables flexible adaptation to graphs of varying sizes and structures across different tasks. Beyond architectural flexibility, we incorporate domain knowledge of information propagation to guide representation learning. Specifically, we define an information propagation graph characterized by binary node states that reflect activation patterns over time. These state transitions are modeled using a kinetic model based on microscopic dynamics \cite{van2008virus,zhang2016dynamics,youssef2011individual}. To integrate this mechanism into training, we further introduce a kinetic-guided objective that aligns the predicted state evolution with the derivatives defined by an ordinary differential equation. This provides physically grounded supervision without requiring labeled states and enhances the ability of our method to learn propagation-aware representations that generalize across heterogeneous graph tasks. Moreover, with the introduction of the kinetic model, the previously mentioned vulnerability to inherent data noise can also be addressed. As a result, the proposed framework is robust and applicable across diverse tasks in a task-specific manner, while also exhibiting strong cross-dataset generalization, making it well-suited for practical use in real-world social media analytics.

Our main contributions are summarized as follows:
\begin{itemize}
    \item We identify a general underlying mechanism across diverse graph-based tasks in social media graph analytics and formalize it as the information propagation graph. This abstraction provides a task-agnostic supervisory signal that enhances representation learning beyond observed features and topologies.
    \item We propose a general architecture that integrates data-driven representation learning with kinetic model-guided supervision. It employs a dual-encoder architecture to capture both structural and contextual semantics, and incorporates a kinetic-guided loss to align learned node state dynamics with physical propagation principles, enabling versatility across heterogeneous graph structures.
    \item We conduct extensive experiments across graph classification, node classification, and link prediction on multiple real-world social media datasets. Our method achieves state-of-the-art performance on each task and demonstrates strong zero- and few-shot transfer capabilities across datasets and task settings.
\end{itemize}

\section{Related Work}
Below we present related work for three supervised learning tasks in social media network analysis, i.e., graph classification, node classification, and link prediction along with their primary applications.

\noindent \textbf{Graph Classification.} The goal of graph classification in social media network is to assign a single label to an entire graph structure. One prominent application is rumor detection, where each instance is represented as a propagation tree or network, and the model must decide if that propagation graph conveys misinformation or factual content. For example, BiGCN treats a rumor propagation tree as a graph and learns bidirectional propagation patterns (top‐down and bottom‐up) to produce a graph‐level embedding for classification \cite{bian2020rumor}. Recent methods further refine the graph‐level representation by incorporating structural priors such as node centrality measures or epidemiological diffusion patterns into the embedding process \cite{cui2024propagation,jiang2025epidemiology}.

\noindent \textbf{Node Classification.} In node classification task, each user (node) in a social network graph is labeled independently, e.g., as “bot” or “human.” BotRGCN applies a relational graph convolutional network (RGCN) over a multi‐relation user graph (retweets, mentions, follows, etc.) to learn per‐node features that capture heterogeneous neighborhood signals, then classifies each node based on its aggregated embedding \cite{feng2021botrgcn}. RGT augments this by adding a self‐attention mechanism over the same multi‐relational graph, allowing the model to weigh information from different relation types dynamically and focus on the most informative neighbors during node embedding construction \cite{feng2022heterogeneity}. 


\noindent \textbf{Link Prediction.} In link prediction, the model receives partial information about an evolving graph (e.g., nodes and edges observed so far in a cascade) and must predict which new user–user or cascade–user edge will form next. DyHGCN tackles this by jointly encoding the static social network and the dynamic diffusion cascade graph with temporal encodings and multi‐head attention; it then scores candidate edges between users and the cascade to forecast the next spread step \cite{yuan2021dyhgcn}. MS‐HGAT extends this by constructing a sequential hypergraph over user interactions within each cascade and applying hypergraph attention layers, supplemented by a memory‐enhanced embedding lookup that captures each user’s evolving preference in the diffusion process \cite{sun2022ms}. RotDiff departs from Euclidean embeddings altogether by mapping users and cascades into a hyperbolic space: rotation transformations in hyperbolic geometry encode hierarchical influence and asymmetric spread patterns, and edge likelihoods are computed via hyperbolic distance metrics \cite{qiao2023rotdiff}. 

Although each family of methods excels within its designated task, they are required significant re-engineering of data representations or architectural components for other tasks. This siloed design incurs heavy modeling overhead when extending to new social media analysis tasks. In contrast, our method unifies all three problem settings under a single backbone with task-specific heads, enabling end-to-end learning for rumor detection, social-bot identification, and diffusion forecasting without rebuilding separate models for each.


\section{Preliminaries}
\subsection{Problem Definition}
In this paper, we aim to design a cross-task architecture that can achieve advantageous performance when trained for each task. As per our discussions earlier, we focus on three key tasks within the domain of supervised social media graph analytics: graph classification, node classification, and link prediction. Note that although the training process is task-specific due to the use of different objectives and hyperparameters, the same model architecture can be directly reused across tasks to minimize engineering costs. Below, we formally define each of these tasks.

\noindent \textbf{Graph Classification:} Graph classification task in the context of social media graph analytics can be formulated as classifying the type of root post using the whole information of a small scale propagation tree. Given a propagation tree ${\mathcal{G}}_{\text{R},i} = (\mathcal{V}_i, \mathcal{E}_i, \boldsymbol{X}_i)$, where $\mathcal{V}_i = \{v_0^{(i)}, v_1^{(i)}, \dots, v_n^{(i)}\}$ denotes the set of nodes corresponding to posts within an event with the root post $v_0$, $\mathcal{E}_i$ denotes the set of edges capturing the reply or retweet structure, and $\boldsymbol{X}_i = [x_0, x_1, \ldots, x_n]^T$ denotes the node feature matrix. The objective of this task is to map each root post $v_0^{(i)}$, along with its associated graph ${\mathcal{G}}_{\text{R},i}$ and textual features $\boldsymbol{X}_i$, to a binary label $y_i \in \{0, 1\}$.

\noindent \textbf{Node Classification:} Node classification is formulated as a supervised binary classification of nodes on a multi-relational social network. The social network is defined as $\mathcal{G}_\text{S} = (\mathcal{V}, \mathcal{E}, \boldsymbol{X})$, where $\mathcal{V} = \{v_i \mid i = 1, 2, \dots, n\}$ denotes the set of nodes, $\mathcal{E} = \cup_{r=1}^R \mathcal{E}r$ is the set of edges under $R$ relation types (e.g. following, follower and reply), and $\boldsymbol{X}$ is the feature matrix, with each row $\boldsymbol{X}_i$ representing the feature vector of node $v_i$. The objective of this task is to use $\mathcal{G}$ and the labels of training nodes $\boldsymbol{Y}_{\text{train}}$ to predict the labels of unseen nodes $\boldsymbol{Y}_{\text{test}}$.

\noindent \textbf{Link Prediction:}
We formulate the link prediction task in the context of social network and its information diffusion cascade sequences, where the goal is to predict the next user likely to be linked in an ongoing diffusion cascade sequence $C_c=\{(u^c_1,t^c_1), \dots, (u^c_n,t^c_n)\}$. Given a dataset $\mathcal{D}$ containing a set of historical cascades $\mathcal{C} = \{ C_1, \cdots, C_{|\mathcal{C}|} \}$ and a social network $\mathcal{G}_{\text{D}} = (\mathcal{V}, \mathcal{E})$ encompassing all users in $\mathcal{D}$, the task is to predict the next link from the current cascade to a potential target user. Specifically, given a query sequence $\mathbf{q} = \{ u_1^q, \cdots, u_m^q \}$, the objective is to predict the next user $u_{m+1}^q$ who will be linked, by leveraging the historical cascade patterns and the structural information in $\mathcal{G}_\text{D}$.

\subsection{Information Propagation Graph}
\label{sec:information_propagation_graph}
\begin{figure}[h]
      \centering
      \includegraphics[width=.9\linewidth]{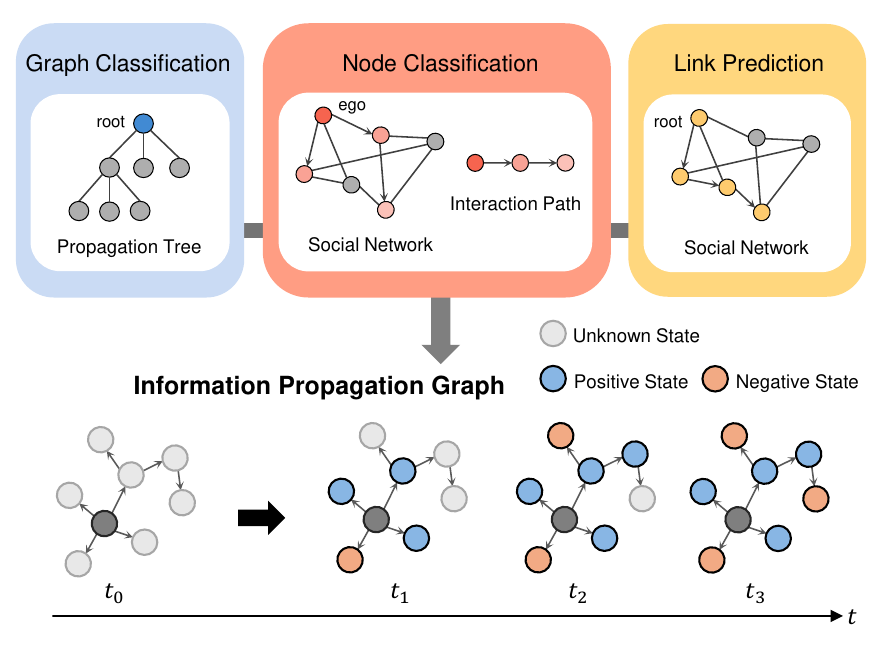}
      \caption{Information Propagation Graph. This graph is an abstraction derived from the propagation tree or the social networks. All the nodes remain in the initial unknown state at time $t_0$. Gradually, nodes at hop-$i$ transition from the unknown state to either a positive or negative state at time step $t_i$.}
      \label{fig:information_propagation_graph}
      \vspace{-3mm}
 \end{figure}

All three tasks in social media networks rely on graph-structured data constructed from user-generated content and interactions. 
Since both information propagation tree and social networks reflect how information spreads among users \cite{hajli2022social, himelein2021bots}, the underlying propagation structure in these tasks can be regarded as information propagation graph, which is denoted as $\mathcal{G}_\text{IP} = (\mathcal{V}, \mathcal{E}, \boldsymbol{S})$ and shown in \Cref{fig:information_propagation_graph}. 
Here, $\mathcal{V}$ and $\mathcal{E}$ represent the sets of nodes and edges, respectively, and $\boldsymbol{S} = \{ U, I_1, I_2 \}$ defines the possible states of each node: an initial unknown state $U$, and two binary informative states, $I_1$ and $I_2$, which abstract states as positive or negative.

In $\mathcal{G}_\text{IP}$, nodes transition from the unknown state $U$ to either $I_1$ or $I_2$ as information propagates through the network, typically emanating from an ego (or root) node. Specifically, at time $t_0$, all nodes are initialized in state $U$, and at each subsequent time step $t_i$, hop-$i$ neighbors of the ego node may transition to either $I_1$ or $I_2$, depending on the dynamics of propagation.
The specific semantics of the binary states depend on the task: they represent supporting and denying comments based on the stances of users, or interacted and non-interacted users (e.g., following, follower, and reply), or active and inactive users in an information cascade session.

\begin{figure*}[h]
      \centering
      \includegraphics[width=.95\linewidth]{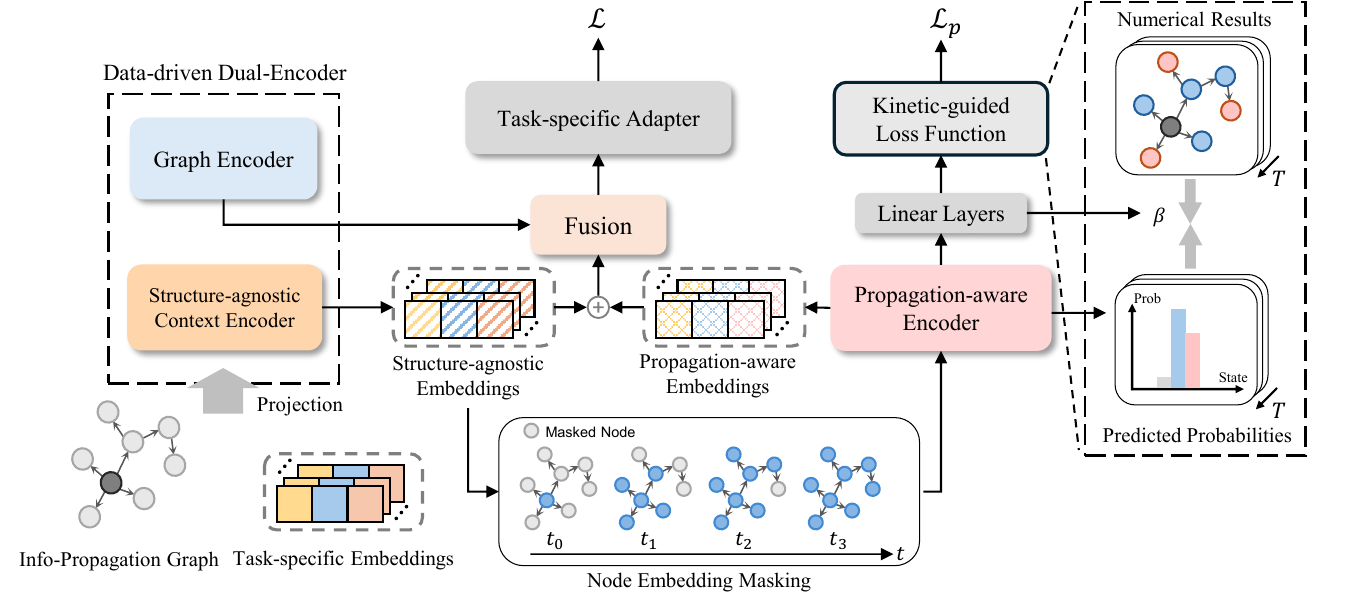}
      \caption{Overview of Robust Propagation-aware Representation Learning Framework}
      \label{fig:framework}
      \vspace{-3mm}
 \end{figure*}

\section{Proposed Framework}

In this section, we present the proposed Robust Propagation-aware Representation Learning (RPRL) framework, a cross-task architecture designed to address graph classification, node classification, and link prediction in social media analytics. As illustrated in \Cref{fig:framework}, RPRL consists of three key components: (1) a data-driven dual-encoder module, (2) a propagation-aware representation learning module, and (3) task-specific adapter and optimization. The details of each module are described below.

\subsection{Data-driven Dual-Encoder}
The variation in graph structures and scales across tasks leads to multifaceted modeling priorities. For instance, propagation trees in graph classification task are typically small in scale, whereas the social networks in node classification and link prediction are significantly larger. Therefore, the emphasis on capturing local structure versus global contextual information varies across tasks, depending on the nature and scale of the underlying graphs. To ensure social graph information is captured at different granularities, we design a parallel architecture that captures both graph topology semantics and global contextual representations for data-driven representation learning. The architecture consists of two key encoders: a Structure-agnostic Context Encoder, which models global contextual relationships by enabling interactions between any pair of nodes regardless of graph connectivity; and a Graph Encoder, which captures local structural dependencies by leveraging the actual graph topology. We first pass the node feature $\boldsymbol{x}_i$ of node $i$ through a task-specific projection layer:
\begin{equation}
    \boldsymbol{\Tilde{x}}_i = \text{projection}(\boldsymbol{x}_i),
    \label{eq:projection}
\end{equation}
where the projection layer is implemented as a linear transformation for both graph and node classification tasks, and as an embedding initialization method following the setup in \cite{sun2022ms} for the link prediction task. Given the transformed $\boldsymbol{\Tilde{x}}_i$ and the information propagation graph  $\mathcal{G}_\text{IP}$, the Structure-agnostic Context Encoder and Graph Encoder can be formulated as:
\begin{equation}
\begin{aligned}
    \boldsymbol{h}_i &= \text{Encoder}_{\text{context}}(\boldsymbol{\Tilde{x}}_i),\\
        \boldsymbol{e}_i &= \text{Encoder}_{\text{graph}}(\mathcal{G}_\text{IP}, \boldsymbol{\Tilde{x}}_i),
\end{aligned}
\end{equation}
where $\boldsymbol{h}_i$ is the output embedding from Structure-agnostic Context Encoder and $\boldsymbol{e}_i$ is the output embedding from Graph Encoder. In this work, we adopt multi-head transformer blocks as the Structure-agnostic Context Encoder and a graph neural network with residual layers based on the setting of \cite{you2020graph} as the Graph Encoder.

\subsection{Propagation-aware Representation Learning}
Most existing solutions on social media graph analytics primarily focus on node-level content features, local neighborhood aggregation, or the temporal dynamics of the data itself. However, they often overlook the structured dynamics of information propagation within the network when guided by kinetic knowledge. As defined in \Cref{sec:information_propagation_graph}, the information propagation graph includes binary propagation states, namely positive and negative, which provide additional semantic signals beyond the observed data. These node-level states enrich the modeling of the propagation process by offering interpretable supervision over how information spreads. The objective of this module is to learn the probability distribution over these binary states from task-specific data, and to incorporate this information into the optimization process.

\begin{algorithm}[t]
\caption{Node Embedding Masking}
\label{alg:masking}
\begin{algorithmic}[1]
 \renewcommand{\algorithmicrequire}{\textbf{Input:}}
 \renewcommand{\algorithmicensure}{\textbf{Output:}}
\REQUIRE Propagation graph $\mathcal{G} = (\mathcal{V}, \mathcal{E})$, node features $\{ h_j \}_{j \in \mathcal{V}}$, ego node $v_{\text{ego}}$, current time step $t$
\ENSURE Masked node embeddings $\{ \tilde{h}_j^{(t)} \}_{j \in \mathcal{V}}$

\FOR{each node $j \in \mathcal{V}$}
    \STATE Compute $d_j \gets \text{dist}(v_{\text{ego}}, j)$
    \IF{$d_j \leq t$}
        \STATE $m_j^{(t)} \gets 1$ 
    \ELSE
        \STATE $m_j^{(t)} \gets 0$ 
    \ENDIF
    \STATE $\tilde{h}_j^{(t)} \gets m_j^{(t)} \cdot h_j$ 
\ENDFOR
\RETURN $\{ \tilde{h}_j^{(t)} \}_{j \in \mathcal{V}}$

\end{algorithmic}
\end{algorithm}
\vspace{-1mm}

\noindent \textbf{Propagation-aware Encoder.} We first introduce a Propagation-aware Encoder that models the simulated propagation process across the graph, enabling the learning of richer representations that better align with real-world propagation behavior. We begin by applying a Node Embedding Masking to simulate the progressive activation of nodes in the propagation process, the implementation is shown in \Cref{alg:masking}. At time step $t$, only nodes that are within a $t$-hop distance from the ego node $v_\text{ego}$ are considered active. These nodes are assumed to have received the propagated information and are allowed to contribute their embeddings to downstream attention computation. In contrast, nodes that lie beyond the $t$-hop boundary are considered inactive at this stage. Instead of directly removing them, we apply zeroing masking to their embeddings, allowing the model to retain a consistent dynamic structure while suppressing their semantic contribution.

Formally, let $d_j = \text{dist}(v_{\text{ego}}, j)$ denote the shortest path distance (in number of hops) from the ego node to node $j$. We define a binary mask $m_j^{(t)} \in \{0, 1\}$ for each node $j$ at time step $t$ as:
\begin{equation}
    m_j^{(t)} =
\begin{cases}
1, & \text{if } d_j \leq t \\
0, & \text{otherwise}.
\end{cases}
\end{equation}

The masked node feature is then computed as:
\begin{equation}
    \boldsymbol{\tilde{h}}_j^{(t)} = m_j^{(t)} \cdot h_j
\end{equation}
where $h_j$ is the structure-agnostic embedding of node $j$, and $\boldsymbol{\tilde{h}}_j^{(t)}$ is its masked embedding at time $t$.

This masking operation is applied prior to the propagation-aware encoder. At each time step, the model processes the masked embeddings of all nodes using shared transformer blocks without graph structure:
\begin{equation}
    \boldsymbol{z}_i^{(t)} = \text{Encoder}_{\text{prop}}(\{\boldsymbol{\tilde{h}}_j^{(t)}\}_{(j,t')\in\mathcal{C}_i^{t}}),
\end{equation}
where 
\begin{equation}
    \mathcal{C}_i^{t} = \{(j,t')|j\in \mathcal{V},0\leq t' \leq t\}.
\end{equation}

Finally, we derive the probabilities of the shared state $s$ for each node:
\begin{equation}
\begin{aligned}
p\left(y_i^{(t)} = s \,\middle|\, \left\{ \boldsymbol{h}_j^{(t)} \right\}_{(j,t') \in \mathcal{C}_i^{(t)}} \right)
&= \mathrm{softmax} \left( \boldsymbol{W}_s \boldsymbol{h}_i^{\mathrm{out}(t)} + \boldsymbol{b}_s \right), \\
&\quad \text{for } s \in \mathcal{S} = \{ U, l_1, l_2\}
\end{aligned}
\label{eq:state_prob}
\end{equation}
where $\boldsymbol{W}_s$ and $b_s$ are learnable parameter and bias, $s$ is the possible state of node $i$. However, relying solely on propagation-aware embedding to enhance data-driven representations is insufficient, as these embeddings are inherently learned through dynamic transformations of the input data. Consequently, their quality remains sensitive to data noise and imperfections, which can lead to suboptimal model performance.

\noindent \textbf{Markov Chain-based Kinetic Model.} To address this challenge, we incorporate domain knowledge to guide the learning of propagation-aware embeddings. The micro-dynamic transitions of node states driven by information propagation in complex networks can be described using a Markov chain-based kinetic model. This model has been widely applied in fields such as epidemiology, information diffusion, and circuit power systems, etc \cite{van2008virus,zhang2016dynamics,youssef2011individual}. Based on the information propagation graph defined in \Cref{sec:information_propagation_graph}, we adapt a binary-state kinetic model that reflects the spread of rumors or information in social networks based on \cite{zhang2016dynamics}. This model captures the fine-grained dynamics by which node states transition from the unknown state to either a positive or negative state during the propagation process. Given the state probabilities $\{U_v(t), I_{1,v}(t), I_{2,v}(t)\}$ of node $i$ at time $t$, the kinetic model is denoted as:
\begin{equation}
\begin{aligned}
\left\{
\begin{aligned}
\frac{dU_v(t)}{dt} &= \sum_k -U_v(t) \beta_k \sum_{j \in \mathbb{N}} a_{v,j} I_{k,v}(t) \\
\frac{dI_{k,v}(t)}{dt} &= U_v(t) \beta_k \sum_{j \in \mathbb{N}} a_{v,j} I_{k,v}(t),
\end{aligned}
\right. \\
\text{s.t. } U_v(t) + \sum_k I_{k,v}(t) &= 1,
\end{aligned}
\end{equation}
where $a_{v,j}$ is the element of adjacency matrix in the information propagation graph, $a_{v,j}=1$ if node $v$ and $j$ are connected, otherwise it is 0, $\beta_k$ is the transition coefficient from unknown state $U$ to $I_k$. 

\noindent \textbf{Kinetic-guided Loss Function.} To ensure that the predictions of state probabilities from \Cref{eq:state_prob} respect these physical constraints, we define a kinetic-guided loss by minimizing the residuals between the predicted state dynamics and the ODE-defined derivatives, enabling physical supervision in the propagation-aware encoder without requiring labeled states. We first denote the predicted probabilities as:
\begin{equation}
\begin{aligned}
\hat{U}_i(t) &= p\left( y_i^{(t)} = U \,\middle|\, \left\{ \tilde{\boldsymbol{h}}_j^{(t)} \right\}_{(j,t') \in \mathcal{C}_i^{(t)}} \right), \\
\hat{I}_{k,i}(t) &= p\left( y_i^{(t)} = I_k \,\middle|\, \left\{ \tilde{\boldsymbol{h}}_j^{(t)} \right\}_{(j,t') \in \mathcal{C}_i^{(t)}} \right).
\end{aligned}
\end{equation}

Given these predictions of probabilities, we then compute the kinetic-guided loss by minimizing the residuals defined by the kinetic model:
\begin{equation}
\begin{aligned}
L_p = 
&\sum_i \left( 
    \frac{d\hat{U}_i(t)}{dt}
    - \left(
        \sum_k -\hat{U}_i(t)\,\hat{\beta}_k \sum_{j \in \mathbb{N}} a_{i,j} \hat{I}_{k,i}(t)
      \right)
\right)^2 \\
&+ \sum_i \sum_k \left(
    \frac{d\hat{I}_{k,i}(t)}{dt}
    - \hat{U}_i(t)\, \hat{\beta}_k \sum_{j \in \mathbb{N}} a_{i,j} \hat{I}_{k,i}(t)
\right)^{2},
\end{aligned}
\end{equation}
where $\hat{\beta}_k$ can be learned by a linear layer with propagation-aware embedding $\boldsymbol{z}_i^{\text{out}(t)}$. We approximate the time derivative $\frac{d\hat{X}_i(t)}{dt}$ using a discretized forward difference quotient. Specifically, given the predicted variable $\hat{X}_i(t)$ at discrete time steps, the derivative is approximated as:
\begin{equation}
    \frac{d\hat{X}_i(t)}{dt} \approx \frac{\hat{X}_i(t + \Delta t) - \hat{X}_i(t)}{\Delta t}.
\end{equation}

In our setting, each time step $t$ corresponds to a discrete hop in the information propagation process, where nodes progressively become activated based on their hop distance from the ego node. Since the propagation steps are uniformly defined over hop distances, we assume a unit time interval ($\Delta t = 1$) between consecutive steps. The forward difference quotient provides a first-order estimate of the temporal change in predicted state probabilities, aligned with the discrete nature of the propagation-aware encoding process.

\subsection{Task-specific Adapters and Model Optimization}
The propagation-aware embedding $\boldsymbol{z}_i^{(t)}$ optimized by the kinetic-guided loss not only captures the dynamic patterns of simulated information propagation, but also benefits from the guidance and constraints provided by domain knowledge. This results in an enhanced structure-agnostic embedding $\hat{\boldsymbol{h}}_i$ that can be effectively integrated with the structure-agnostic embedding $\boldsymbol{h}_i$ as follows:
\begin{equation}
    \hat{\boldsymbol{h}}_i = \boldsymbol{h}_i + \sigma(\boldsymbol{z}_i^{(t)}),
\end{equation}
where $\sigma$ is an activation function. Since the graph structures and node features vary significantly across tasks, and each task may emphasize different aspects of the data, it is necessary to fuse the enhanced structure-agnostic embedding with the graph embedding. This fusion allows the model to simultaneously capture information from both global and structural perspectives, thereby improving the overall representation capacity and robustness of our method. We then input the graph embedding $\boldsymbol{e}_i$ and enhanced structure-agnostic embedding $\hat{\boldsymbol{h}}_i$ to a fusion layer:
\begin{equation}
    \boldsymbol{o}_i = \gamma\cdot\boldsymbol{e}_i + (1-\gamma)\cdot\hat{\boldsymbol{h}}_i,
\end{equation}
where $\gamma$ is the coefficient that controls the contribution of graph structural information during learning. The output embedding $\boldsymbol{o}_i$ is passed through a task-specific adapter to produce the final prediction.

\noindent \textbf{Graph Classification.} For graph classification, each input sample consists of a small-scale propagation tree $\mathcal{G}_{\text{R}}$ along with its corresponding node features $\boldsymbol{X}_i$. We input $\boldsymbol{X}_i$ to \Cref{eq:projection} and obtain the output embeddings $\boldsymbol{o}_i$ through the above learning process, the graph classification adapter first applies a graph pooling operation, followed by a linear transformation layer to produce the final prediction $p_g$:
\begin{equation}
\begin{aligned}
       \boldsymbol{o}_g &= \text{Pooling}(\boldsymbol{O}),\\
        p_g &= \text{softmax}(\boldsymbol{W}_g \boldsymbol{o}_g + b_g),
\end{aligned}
\end{equation}
where $\boldsymbol{O}$ is the combination of all nodes' output embeddings, $\boldsymbol{W}_g$ and $b_g$ are the parameters of graph-level output layer. 

\noindent \textbf{Node Classification.} For node classification, the input consists of a large-scale social network $\mathcal{G}_{\text{S}}$ along with node features $\boldsymbol{X}$, where each node serves as an individual classification target. We input $\boldsymbol{X}$ to \Cref{eq:projection} and learn the output embedding $\boldsymbol{o}_i$ of each node through the preceding steps, the final prediction for each node is obtained by applying a linear transformation:
\begin{equation}
    p_i = \text{softmax}(\boldsymbol{W}_o \boldsymbol{o}_i + b_o),
\end{equation}
where $\boldsymbol{W}_o$ and $b_o$ are the parameters of node-level output layer. 

\noindent \textbf{Link Prediction.} For link prediction, each input sample consists of a cascade sequence $C_c$ and a large-scale social network $\mathcal{G}_{\text{D}}$. We input the concatenated feature $\boldsymbol{X}_{\text{all}} = \text{concatenate}(\boldsymbol{x}_1, \dots, \boldsymbol{x}_N)$ to \Cref{eq:projection}, and derive the output embedding  $\boldsymbol{O}_{\text{all}}$ from the previous steps, to predict which node is most likely to be linked to $C_c$. A linear output layer is applied to perform this prediction:
\begin{equation}
    \boldsymbol{P} = \text{softmax}(\boldsymbol{W}_p \boldsymbol{O}_{\text{all}} + \text{Mask}_p),
\end{equation}
where $\boldsymbol{W}_p$ is the parameter matrix of link prediction output layer. $\boldsymbol{P}$ is the matrix of activation probabilities for all users. $\text{Mask}_p$ is using for masking the users who have been in the cascade. $\text{Mask}_{i,p}=-\infty$ if user $i$ is already in the cascade, else $\text{Mask}_{i,p} = 0$.

\noindent \textbf{Model Optimization.} Finally, we derive three types of predictions corresponding to different tasks: (1) the graph-level prediction $p_g$ for graph classification, where the goal is to predict the category of the root post in a propagation tree; (2) the node-level prediction $p_i$ for node classification, which assigns category labels to each node in a social network; and (3) the link-level prediction $\boldsymbol{P}$ for link prediction, which predicts the likelihood of each node in the social network being linked to a given cascade sequence $C_c$. For all the tasks, a softmax function is applied to the model outputs to obtain probability distributions. The resulting predictions are then supervised using a cross-entropy loss, denoted as $\mathcal{L}_s$, and the final optimization objective function of RPRL is:
\begin{equation}
    \mathcal{L} = \mathcal{L}_s + \lambda\mathcal{L}_p,
\end{equation}
where $\mathcal{L}_p$ is the kinetic-guided loss function, $\lambda$ controls the contribution of the kinetic-guided loss in the overall objective.

\section{Experiments}
\subsection{Experimental Settings}
In our experiments, we evaluate the effectiveness of the proposed RPRL on three real-world applications: rumor detection, social-bot detection, and information diffusion. These applications correspond to the tasks of graph classification, node classification, and link prediction, respectively.

\noindent \textbf{Datasets:} We evaluate RPRL on six datasets covering all three tasks. For rumor detection, we use DRWeibo  \cite{cui2024propagation} and Weibo \cite{ma2016detecting}; for social-bot detection, we use Twibot-22 \cite{feng2022twibot} and MGTAB \cite{shi2023mgtab}; and for information diffusion prediction, we use the Christianity \cite{sankar2020inf} and Twitter \cite{hodas2014simple} datasets. In the rumor detection setting, following \cite{cui2024propagation, jiang2025epidemiology}, we represent node-level textual content using Word2Vec embeddings. For social-bot detection, we follow the setting of \cite{zhou2023detecting}, sampling 10,000 nodes from Twibot-22 while preserving the original label distribution. For information diffusion prediction, we adopt the initialization strategy from \cite{sun2022ms}, where node embeddings are initialized using graph convolutional operations. To ensure fair comparison, we follow the evaluation protocols used by the baselines in each task \cite{jiang2025epidemiology, feng2022heterogeneity, wang2024information}, and split each dataset into training, validation, and test sets using a 6:2:2 ratio for rumor detection, 7:1:2 for social-bot detection, and 8:1:1 for information diffusion prediction.

\noindent \textbf{Baselines:} We compare RPRL with 6 widely used and state-of-the-art baselines for each of the three tasks, respectively.
\begin{itemize}
    \item Rumor Detection: Our baselines include graph-based rumor detection methods ResGCN \cite{you2020graph}, KAGIN \cite{bresson2024kagnns}, GACL \cite{sun2022rumor}, GARD \cite{tao2024semantic}, RAGCL \cite{cui2024propagation}, EIN \cite{jiang2025epidemiology}.
    \item Social-bot Detection: Our baselines include  GAT, BotRGCN \cite{feng2021botrgcn}, SHGN \cite{lv2021we}, RGT \cite{feng2022heterogeneity}, BotMoE \cite{liu2023botmoe}, SEBot \cite{yang2024sebot}, which are state-of-the-art methods based on graph neural networks.
    \item Information Diffusion Prediction: Our baselines include both dynamic and hypergraph methods like FOREST \cite{yang2021full}, DyHGCN \cite{yuan2021dyhgcn}, MSHGAT \cite{sun2022ms}, RotDiff\cite{qiao2023rotdiff}, MINDS \cite{jiao2024enhancing}, GODEN \cite{wang2024information}.
\end{itemize}

\noindent \textbf{Evaluation Metrics and Implementation Details:} 
We adopt standard evaluation metrics tailored to each task. For the rumor detection task, we use Accuracy (Acc.), ROC-AUC (AUC), and F1-score (F1). For social-bot detection, the evaluation metrics include Accuracy (Acc.), Balanced Accuracy (B-Acc.), and F1-score (F1). For the information diffusion prediction task, we report Hit@K and Mean Average Precision at K (MAP@K), where $K \in \{10, 100\}$.

 We implement the proposed RPRL framework using PyTorch and conduct all experiments on an NVIDIA RTX 4090 GPU. Detailed parameter settings for RPRL across all tasks are available in the released code repository. In addition, we provide an analysis of the impact of the hyperparameters $\gamma$ and $\lambda$ in \Cref{sec:hyperparam_analyse}.

\begin{table}[ht]
\centering
\caption{Overall comparison results for the graph-based rumor detection task (\%).}
\label{tab:overall_rumor}
\renewcommand{\arraystretch}{1}
\begin{tabular}{llccc}
\toprule
Dataset   & Model      & Acc.     & AUC           & F1             \\
\midrule
          & ResGCN     & 87.79$\pm$1.29     & 87.73$\pm$1.14     & 86.98$\pm$1.09     \\
          & KAGIN      & 86.75$\pm$0.29     & 86.54$\pm$0.64     & 85.61$\pm$1.42     \\
          & GACL       & 83.14$\pm$3.17     & 83.51$\pm$3.01     & 84.15$\pm$2.38     \\
          & GARD       & 88.58$\pm$0.76     & 88.55$\pm$0.77     & 87.95$\pm$0.66     \\
DRWeibo   & RAGCL      & 88.50$\pm$0.53     & 88.37$\pm$0.62     & 87.57$\pm$0.72     \\
          & EIN        & {89.04$\pm$0.47} & {88.95$\pm$0.53} & {88.28$\pm$0.53} \\
          & RPRL w/o pt        & \underline{89.49$\pm$0.29} & \underline{89.47$\pm$0.13} & \underline{88.86$\pm$0.07} \\
          \cmidrule(lr){2-5}
          & RAGCL-pt   & 88.83$\pm$0.53     & 88.63$\pm$0.54     & 87.77$\pm$0.59     \\
          & EIN-pt     & 88.54$\pm$0.47     & 88.57$\pm$0.34     & 88.02$\pm$0.37     \\
          & RPRL (ours)        & \textbf{89.53$\pm$0.59} & \textbf{89.61$\pm$0.49} & \textbf{89.18$\pm$0.59} \\
\midrule
          & ResGCN     & 93.89$\pm$1.67     & 93.92$\pm$1.63     & 93.82$\pm$1.85     \\
          & KAGIN      & 92.44$\pm$0.68     & 92.46$\pm$0.70     & 92.28$\pm$0.86     \\
          & GACL       & 94.37$\pm$0.83     & 94.39$\pm$0.82     & 94.36$\pm$0.90     \\
          & GARD       & 94.43$\pm$0.45     & 94.41$\pm$0.43     & 94.56$\pm$0.56     \\
Weibo     & RAGCL      & 93.94$\pm$0.68     & 93.95$\pm$0.69     & 93.87$\pm$0.82     \\
          & EIN        & 95.39$\pm$0.61     & 95.38$\pm$0.59     & 95.45$\pm$0.69     \\
          & RPRL w/o pt        & {95.57$\pm$0.15} & {95.71$\pm$0.15} & \underline{95.75$\pm$0.12} \\
          \cmidrule(lr){2-5}
          & RAGCL-pt   & 94.59$\pm$0.23     & 94.59$\pm$0.24     & 94.58$\pm$0.26     \\
          & EIN-pt     & \underline{95.66$\pm$0.38} & \underline{95.66$\pm$0.38} & {95.70$\pm$0.32} \\
          & RPRL (ours)        & \textbf{96.20$\pm$0.99} & \textbf{96.19$\pm$0.98} & \textbf{96.25$\pm$0.98} \\
\bottomrule
\end{tabular}
\vspace{-3mm}
\end{table}

\begin{table}[ht]
\centering
\caption{Overall comparison results for the social-bot detection task (\%).}
\label{tab:overall_socialbot}
\renewcommand{\arraystretch}{1}
\begin{tabular}{llccc}
\toprule
Dataset    & Model        & Acc.      & B-Acc.           & F1            \\
\midrule
          & BotRGCN      & 86.93$\pm$0.37     & {69.46$\pm$0.14} & {50.12$\pm$0.22} \\
           & GAT          & 87.20$\pm$0.21     & 67.30$\pm$0.24     & 47.30$\pm$0.60     \\
           & SHGN         & 87.23$\pm$0.43     & 67.49$\pm$2.77     & 47.48$\pm$4.68     \\
Twibot-22           & RGT          & 87.32$\pm$0.33     & 69.13$\pm$2.17     & 49.99$\pm$2.65     \\
           & BotMoE       & 87.10$\pm$0.23     & 48.14$\pm$5.83     & 49.58$\pm$0.40     \\
           & RPRL w/o pt           & {86.87$\pm$0.48} & \underline{71.44$\pm$0.28} & \underline{52.57$\pm$0.24} \\
           \cmidrule(lr){2-5}
           & BotRGCN-pt   & 86.97$\pm$0.14     & 66.32$\pm$3.88     & 45.14$\pm$6.17     \\
           & RGT-pt       & \textbf{87.43$\pm$0.14} & 66.53$\pm$1.40     & 46.28$\pm$2.43     \\
           & RPRL (ours)          & \underline{87.42$\pm$0.07} & \textbf{72.61$\pm$0.36} & \textbf{54.64$\pm$0.32} \\
\midrule
      & BotRGCN      & 89.13$\pm$2.35     & 83.60$\pm$5.58     & 77.86$\pm$6.88     \\
           & GAT          & 87.66$\pm$1.25     & 84.70$\pm$0.05     & 77.57$\pm$1.02     \\
           & SHGN         & 87.66$\pm$3.74     & 86.46$\pm$2.31     & 78.88$\pm$5.09     \\
           & RGT          & {89.88$\pm$0.94} & 86.29$\pm$2.09     & 80.69$\pm$2.45     \\
 MGTAB          & BotMoE       & 88.69$\pm$0.64     & 86.71$\pm$1.17     & 80.29$\pm$0.69     \\
           & SEBot        & 89.41$\pm$1.86     & \underline{86.95$\pm$0.99} & 80.64$\pm$2.28     \\
           & RPRL w/o pt          & \underline{90.21$\pm$1.25} & {86.17$\pm$2.24} & \underline{81.59$\pm$2.89} \\
           \cmidrule(lr){2-5}
           & BotRGCN-pt   & 89.72$\pm$0.28     & 86.46$\pm$0.36     & {80.77$\pm$0.67} \\
           & RGT-pt       & 88.10$\pm$0.21     & 82.59$\pm$0.12     & 76.34$\pm$0.45     \\
           & RPRL (ours)          & \textbf{90.45$\pm$1.04} & \textbf{87.35$\pm$2.39} & \textbf{82.07$\pm$2.71} \\
\bottomrule
\end{tabular}
\vspace{-3mm}
\end{table}

\begin{table}[ht]
\centering
\caption{Overall comparison results for the information diffusion prediction task (\%).}
\label{tab:overall_idp}
\setlength{\tabcolsep}{2pt}
\renewcommand{\arraystretch}{1}
\begin{tabular}{llcccc}
\toprule
Dataset       & Model        & Hits@10         & MAP@10          & Hits@100        & MAP@100         \\
\midrule
  & FOREST       & 23.88$\pm$1.90       & 16.26$\pm$0.47       & 56.80$\pm$2.37       & 17.33$\pm$0.42       \\
              & DyHGCN       & 30.13$\pm$0.01       & 18.69$\pm$0.39       & 60.72$\pm$0.63       & 19.74$\pm$0.41       \\
              & MSHGAT       & 30.47$\pm$1.74       & 17.24$\pm$0.86       & 61.05$\pm$0.79       & 18.34$\pm$0.72       \\
              & RotDiff      & {30.92$\pm$0.16} & 19.14$\pm$0.70       & \textbf{63.72$\pm$0.47} & 20.33$\pm$0.81       \\
  Chris.            & MINDS        & 27.79$\pm$1.10       & 17.88$\pm$0.30       & 57.59$\pm$1.58       & 18.94$\pm$0.44       \\
              & GODEN        & 30.25$\pm$2.57       & \underline{19.88$\pm$0.75} & 58.93$\pm$2.68       & \underline{20.84$\pm$0.69} \\
              & RPRL w/o pt & \underline{32.03$\pm$0.16}  & 18.66$\pm$0.19 & 62.16$\pm$0.16 & 19.76$\pm$0.15 \\
              \cmidrule(lr){2-6}
              & RotDiff-pt   & 30.81$\pm$0.32       & 19.20$\pm$0.60       & 62.16$\pm$3.01       & 20.24$\pm$0.69       \\
              & GODEN-pt     & 28.79$\pm$0.22       & 18.99$\pm$0.26       & 57.48$\pm$0.33       & 19.95$\pm$0.25       \\
              & RPRL (ours)        & \textbf{33.70$\pm$1.27} & \textbf{20.30$\pm$0.13} & \underline{62.95$\pm$0.63} & \textbf{21.32$\pm$0.06} \\
\midrule
       & FOREST       & 23.33$\pm$0.04       & 16.47$\pm$0.28       & 42.55$\pm$0.44       & 17.05$\pm$0.26       \\
              & DyHGCN       & 33.79$\pm$0.03       & 22.80$\pm$0.15       & 57.28$\pm$0.16       & 23.61$\pm$0.16       \\
              & MSHGAT       & 30.61$\pm$0.37       & 19.76$\pm$0.11       & 56.41$\pm$0.49       & 20.63$\pm$0.14       \\
              & RotDiff      & 35.77$\pm$0.16       & 23.07$\pm$0.20       & 62.22$\pm$0.23       & 23.95$\pm$0.21       \\
   Twitter           & MINDS        & 28.67$\pm$1.44       & 15.89$\pm$1.13       & 53.50$\pm$1.22       & 16.73$\pm$1.13       \\
              & GODEN        & \underline{37.78$\pm$0.04} & {24.45$\pm$0.09} & \textbf{64.80$\pm$0.25} & {25.40$\pm$0.08} \\
               & RPRL w/o pt      & {37.43$\pm$0.35} & \underline{24.94$\pm$0.21} & \underline{63.89$\pm$0.06} & \underline{25.86$\pm$0.19} \\
              \cmidrule(lr){2-6}
              & RotDiff-pt   & 34.92$\pm$0.48       & 22.56$\pm$0.70       & 62.10$\pm$0.59       & 23.49$\pm$0.69       \\
              & GODEN-pt     & 34.31$\pm$0.44       & 21.22$\pm$0.48       & 62.65$\pm$1.77       & 22.19$\pm$0.43       \\
              & RPRL (ours)          & \textbf{38.05$\pm$0.14} & \textbf{25.29$\pm$0.07} & {63.60$\pm$0.21} & \textbf{26.18$\pm$0.08} \\
\bottomrule
\end{tabular}

\end{table}

\subsection{Overall Comparison}
\label{sec:overall}
\noindent \textbf{Setup:} For all baseline models across all tasks, we adopt end-to-end training. The proposed RPRL framework follows a cross-dataset pre-training and fine-tuning setting. For instance, in the rumor detection task, if Weibo is the target dataset, RPRL is first pre-trained on the DRWeibo dataset, and then fine-tuned and evaluated on Weibo. To ensure a fair comparison, we apply the same cross-dataset pre-training strategy to the top two best-performing baseline models, which are denoted with the suffix “-pt”. We also compare the version of RPRL without pre-training with other baselines, which is denoted as RPRL w/o pt.

\noindent \textbf{Result:} We compare the proposed RPRL framework against baseline models across all three tasks, with the results summarized in \Cref{tab:overall_rumor,tab:overall_socialbot,tab:overall_idp}. The best results are highlighted in bold, and the second-best results are underlined. Based on these results, we make the following observations: (1) Our method achieves state-of-the-art performance across all three tasks. Even without pre-training, the RPRL model outperforms most baseline methods in the majority of cases. This demonstrates that our approach not only supports cross-task generalization under the same architecture, but also exhibits strong cross-dataset transferability and robustness to data variations. (2) For those baseline models that are also pre-trained across datasets, their performance is unstable and sometimes even worse than their non-pretrained counterparts (e.g., RAGCL and EIN on DRWeibo). This phenomenon suggests that these models have limited knowledge transfer capability and struggle to generalize to unseen data distributions. In contrast, RPRL consistently maintains high performance even under domain shift, highlighting its superior adaptability and robustness.

\subsection{Zero- and Few-shot Capability of RPRL}
\noindent \textbf{Setup:} To further evaluate the cross-dataset generalization and data efficiency of RPRL, we conduct zero- and few-shot experiments across all three tasks. We adopt the cross-dataset pre-train and fine-tune setting in \Cref{sec:overall} to pre-train all the methods and fine-tune on the target dataset. For the few-shot experiments, we randomly select 1\% and 5\% of the target dataset for the one-shot and five-shot settings, respectively. In the zero-shot setting, we directly evaluate performance on the target dataset without any fine-tuning.

\begin{figure*}[ht]
    \centering
    \begin{subfigure}[t]{0.325\textwidth}
        \centering
        \includegraphics[width=\linewidth]{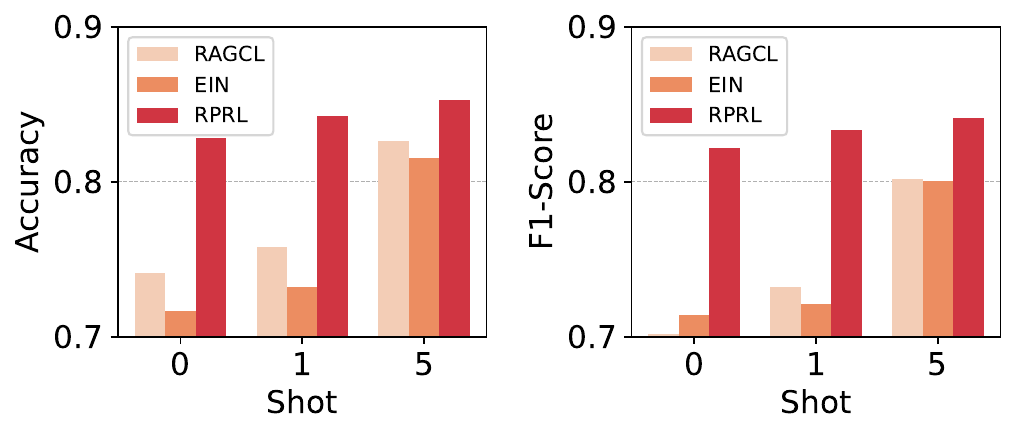}
        \caption{DRWeibo}
    \end{subfigure}
    \hfill
    \begin{subfigure}[t]{0.325\textwidth}
        \centering
        \includegraphics[width=\linewidth]{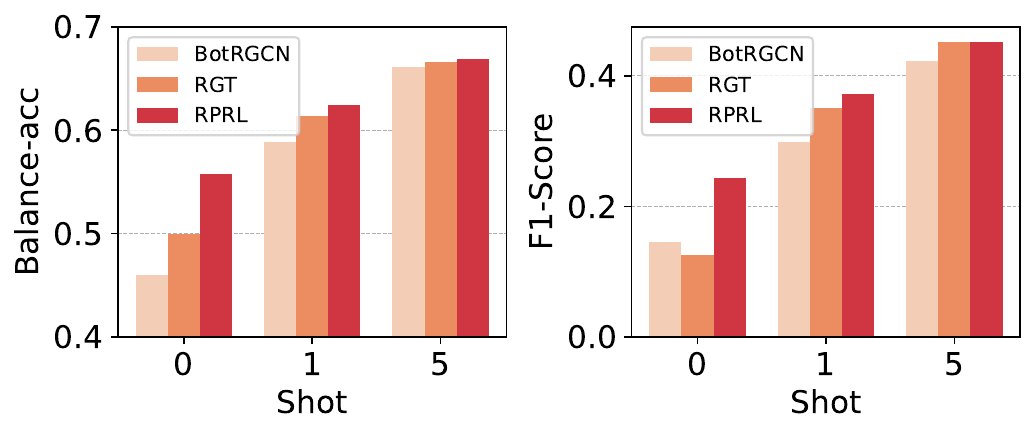}
        \caption{Twibot-22}
    \end{subfigure}
    \hfill
    \begin{subfigure}[t]{0.325\textwidth}
        \centering
        \includegraphics[width=\linewidth]{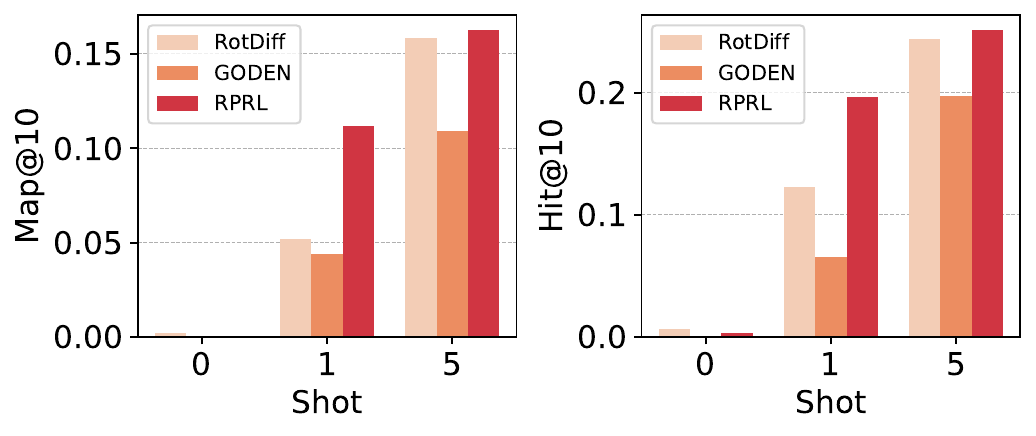}
        \caption{Christianity}
    \end{subfigure}

    \vspace{0.5em} 

    \begin{subfigure}[t]{0.325\textwidth}
        \centering
        \includegraphics[width=\linewidth]{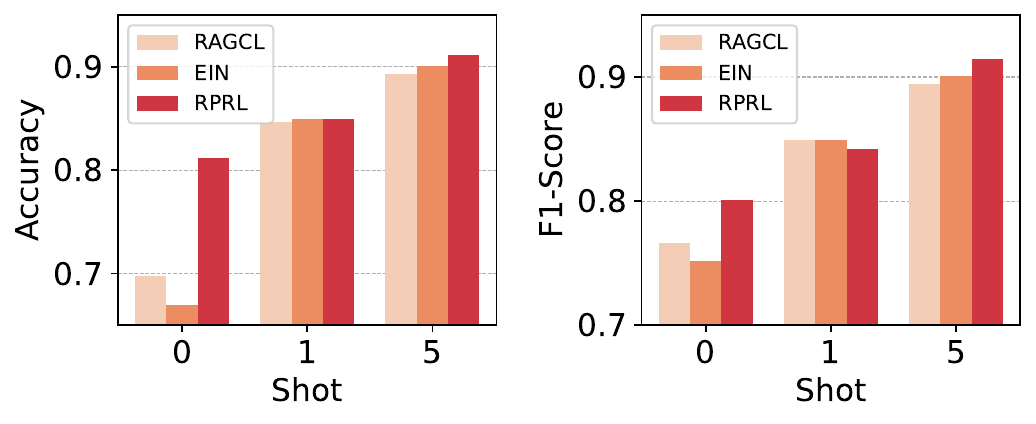}
        \caption{Weibo}
    \end{subfigure}
    \hfill
    \begin{subfigure}[t]{0.325\textwidth}
        \centering
        \includegraphics[width=\linewidth]{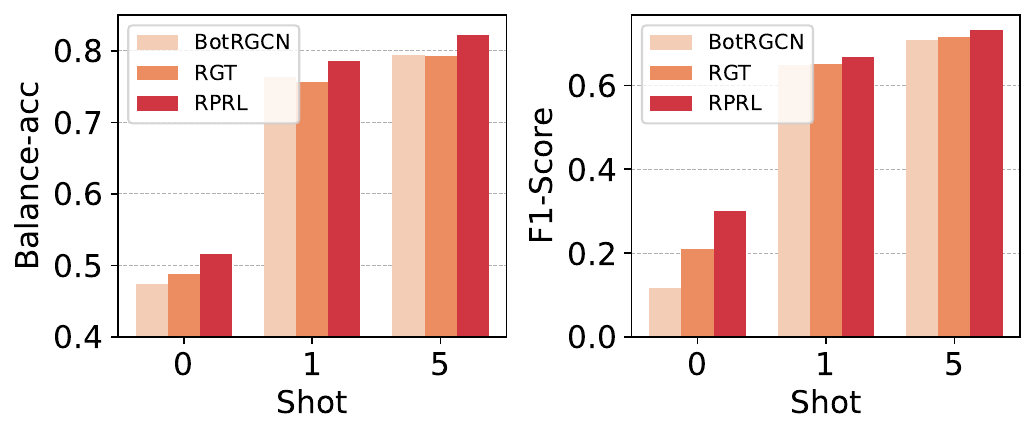}
        \caption{MGTAB}
    \end{subfigure}
    \hfill
    \begin{subfigure}[t]{0.325\textwidth}
        \centering
        \includegraphics[width=\linewidth]{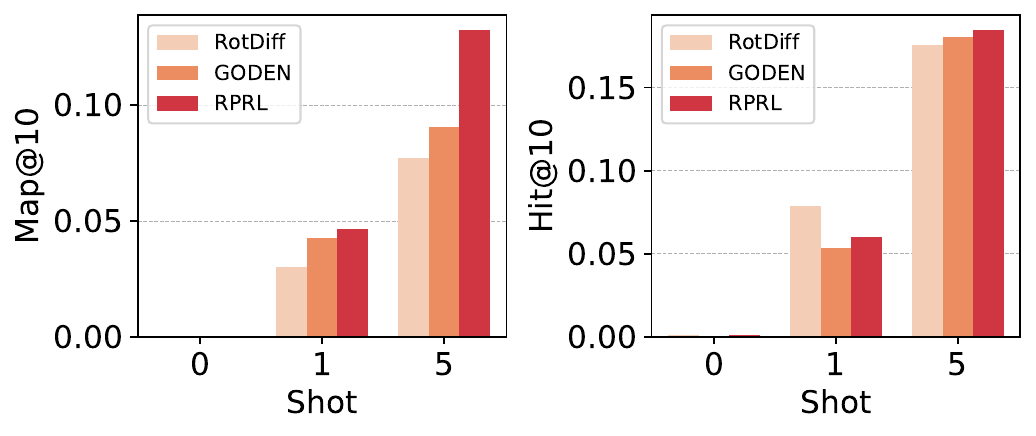}
        \caption{Twitter}
    \end{subfigure}

    \caption{Performance of RPRL and baselines across three tasks under different zero- and few-shot settings.}
    \label{fig:zero_few_shot}
\end{figure*}

\noindent \textbf{Result:} 
As shown in \Cref{fig:zero_few_shot}, RPRL demonstrates strong cross-dataset generalization, consistently outperforming all baseline methods in zero-shot settings in most cases. This trend is especially evident in the rumor detection and social-bot detection tasks, where RPRL achieves superior performance without access to any labeled samples from the target domain. Notably, in the rumor detection task, RPRL’s zero-shot performance even surpasses the five-shot results of the baseline models, underscoring its exceptional cross-domain transferability.

As more supervision becomes available, RPRL continues to show clear performance gains. In both one-shot and five-shot scenarios, its advantages remain consistent across all tasks and evaluation metrics: accuracy and F1-score for rumor detection, balanced accuracy and F1-score for social-bot detection, and MAP@10 and Hit@10 for information diffusion prediction. These results highlight RPRL’s strong data efficiency and adaptability in low-resource settings.

An exception arises in the information diffusion prediction task (Christianity and Twitter datasets), where all methods, including our RPRL, exhibit relatively poor performance in the zero-shot setting. This outcome is expected, as the task is framed as a retrieval-based link prediction problem that involves selecting the next user to be activated from a large candidate pool. Without any labeled data in the target domain, models struggle to transfer the retrieval objective, making zero-shot prediction particularly challenging. Nevertheless, RPRL quickly outperforms all baselines once minimal supervision (e.g., one-shot) is introduced, reaffirming its strong adaptability even in complex task formulations.

\subsection{Ablation Study}

To investigate the contributions of different components in RPRL, we conduct an ablation study across all three tasks, as shown in \Cref{tab:abaltion}. We compare the full model with three variants: (1) RPRL w/o pt, which removes cross-dataset pre-training; (2) RPRL w/o pe, which disables propagation-aware embedding learning by removing the propagation-aware encoder and kinetic-guided loss; and (3) RPRL-rg, which replaces the Markov chain-based kinetic model with a simpler regular kinetic model for computing the kinetic-guided loss.

The results demonstrate several key findings. Above all, removing the propagation-aware embedding module (RPRL w/o pe) leads to a consistent and notable performance drop across all datasets and tasks. This highlights the importance of modeling propagation dynamics via the propagation-aware encoder and kinetic-guided supervision. Besides, the RPRL w/o pt variant generally performs worse than the fully pre-trained version, indicating that cross-dataset pre-training improves generalization and boosts performance. The RPRL-rg variant, which replaces the kinetic-guided loss function with a regular kinetic model, also underperforms compared to the full model, indicating the advantage of our modified Markov chain-based kinetic formulation for guiding information propagation modeling.

\begin{table}[ht]
\centering
\caption{Performance comparison of RPRL variants across three tasks (\%). Best results are in bold.}
\label{tab:abaltion}
\renewcommand{\arraystretch}{1}
\begin{tabular}{lllcc}
\toprule
\textbf{Task} & \textbf{Dataset} & \textbf{Model} & \textbf{Acc.} & \textbf{F1} \\
\midrule

\multirow{8}{*}{RD} 
& DRWeibo 
  & RPRL w/o pt & 89.24$\pm$0.18 & 88.73$\pm$0.11 \\
& & RPRL w/o pe & \textbf{89.58$\pm$0.65} & {88.98$\pm$0.48} \\
& & RPRL-rg & 88.41$\pm$0.41 & 87.57$\pm$0.76 \\
& & RPRL        & {89.53$\pm$0.47} & \textbf{89.18$\pm$0.59} \\
\cmidrule{2-5}
& Weibo
  & RPRL w/o pt & 95.55$\pm$0.83 & 95.62$\pm$0.87 \\
& & RPRL w/o pe & 94.86$\pm$1.36 & 94.90$\pm$1.46 \\
& & RPRL-rg & 95.39$\pm$1.36 & 95.45$\pm$1.65 \\
& & RPRL        & \textbf{96.20$\pm$0.99} & \textbf{96.25$\pm$0.98} \\

\midrule
\textbf{Task} & \textbf{Dataset} & \textbf{Model} & \textbf{B-Acc.} & \textbf{F1} \\
\midrule
\multirow{8}{*}{SBD}
& TwiBot-22
  & RPRL w/o pt & 71.44$\pm$0.28 & 52.57$\pm$0.24 \\
& & RPRL w/o pe & 68.85$\pm$1.14 & 49.97$\pm$1.65 \\
& & RPRL-rg & 70.52$\pm$0.92 & 51.36$\pm$1.34 \\
& & RPRL        & \textbf{72.61$\pm$0.36} & \textbf{54.64$\pm$0.32} \\
\cmidrule{2-5}
& MGTAB
  & RPRL w/o pt & 86.17$\pm$2.24 & 81.09$\pm$2.96 \\
& & RPRL w/o pe & 86.93$\pm$2.64 & 81.59$\pm$2.89 \\
& & RPRL-rg & 86.18$\pm$2.25 & 80.92$\pm$2.72 \\
& & RPRL        & \textbf{87.35$\pm$2.39} & \textbf{82.07$\pm$2.71} \\

\midrule
\textbf{Task} & \textbf{Dataset} & \textbf{Model} & \textbf{Hits@10} & \textbf{MAP@10} \\
\midrule

\multirow{8}{*}{IDP}
& Christianity
  & RPRL w/o pt & 32.03$\pm$0.16 & 18.66$\pm$0.19 \\
& & RPRL w/o pe & 31.92$\pm$0.01 & 19.31$\pm$0.16 \\
& & RPRL-rg & 31.92$\pm$1.27 & 19.62$\pm$0.72 \\
& & RPRL        & \textbf{33.70$\pm$1.27} & \textbf{20.30$\pm$0.13} \\
\cmidrule{2-5}
& Twitter
  & RPRL w/o pt & 37.43$\pm$0.35 & 24.94$\pm$0.21 \\
& & RPRL w/o pe & 37.74$\pm$0.12 & 24.62$\pm$0.54 \\
& & RPRL-rg & 37.28$\pm$0.11 & 24.90$\pm$0.16 \\
& & RPRL        & \textbf{38.05$\pm$0.14} & \textbf{25.29$\pm$0.07} \\

\bottomrule
\end{tabular}
\vspace{-4mm}
\end{table}

\subsection{Hyperparameter Analysis}
\label{sec:hyperparam_analyse}

To analyze the sensitivity of our model to key hyperparameters, we conduct experiments on three representative datasets from each task: DRWeibo (rumor detection), MGTAB (social bot detection), and Christianity (information diffusion prediction). We analyze the influences of two hyperparameters: $\gamma$, which controls the contribution of graph encoder, and $\lambda$, which controls the contribution of the kinetic-guided loss in the overall optimization process.

\begin{figure}[ht]
    \centering

    \begin{subfigure}[t]{0.15\textwidth}
        \centering
        \includegraphics[width=\linewidth]{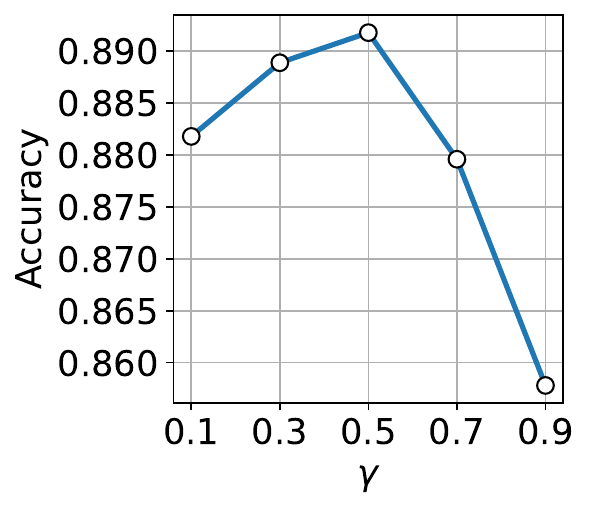}
        \caption{$\gamma$ on DRWeibo.}
    \end{subfigure}
    \begin{subfigure}[t]{0.15\textwidth}
        \centering
        \includegraphics[width=\linewidth]{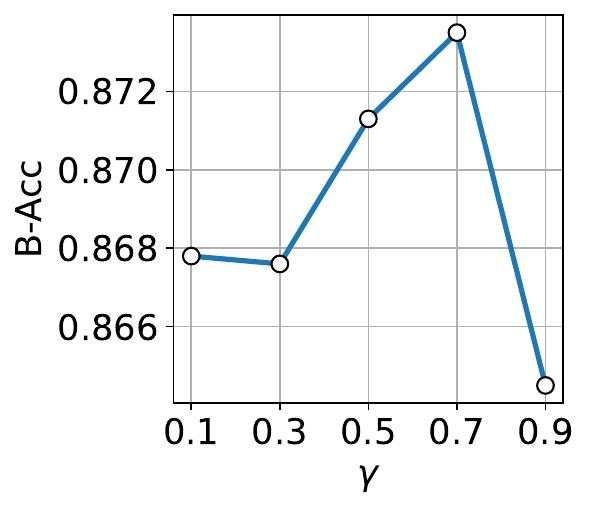}
        \caption{$\gamma$ on MGTAB.}
    \end{subfigure}
    \begin{subfigure}[t]{0.15\textwidth}
        \centering
        \includegraphics[width=\linewidth]{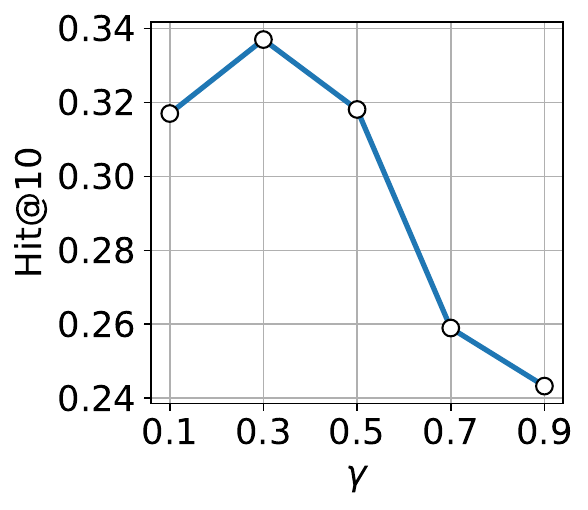}
        \caption{$\gamma$ on Christianity.}
    \end{subfigure}

    \vspace{1em}

    \begin{subfigure}[t]{0.15\textwidth}
        \centering
        \includegraphics[width=\linewidth]{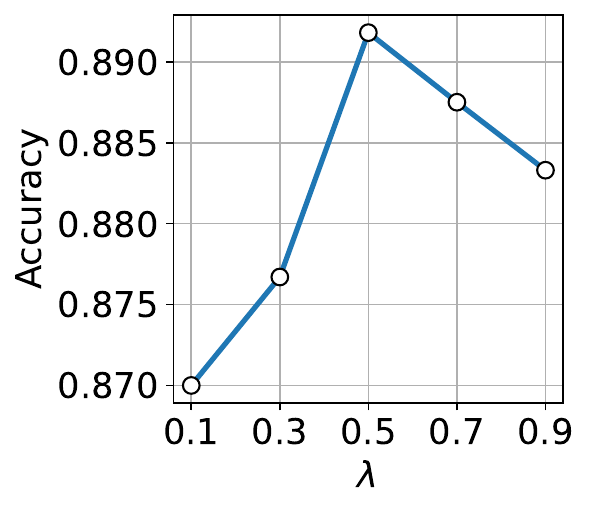}
        \caption{$\lambda$ on DRWeibo.}
    \end{subfigure}
    \begin{subfigure}[t]{0.15\textwidth}
        \centering
        \includegraphics[width=\linewidth]{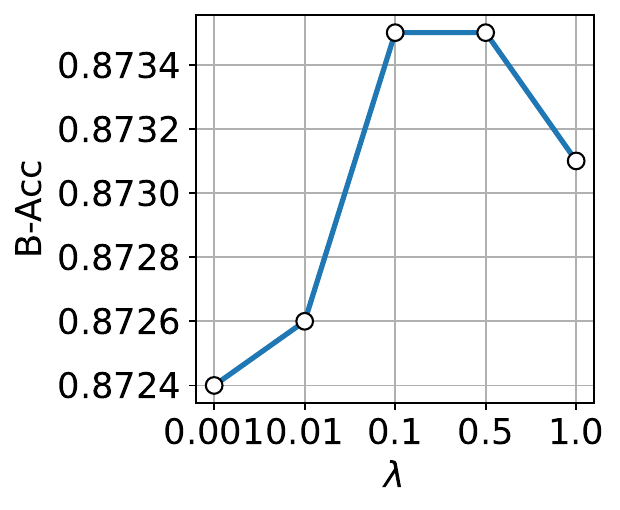}
        \caption{$\lambda$ on MGTAB.}
    \end{subfigure}
    \begin{subfigure}[t]{0.15\textwidth}
        \centering
        \includegraphics[width=\linewidth]{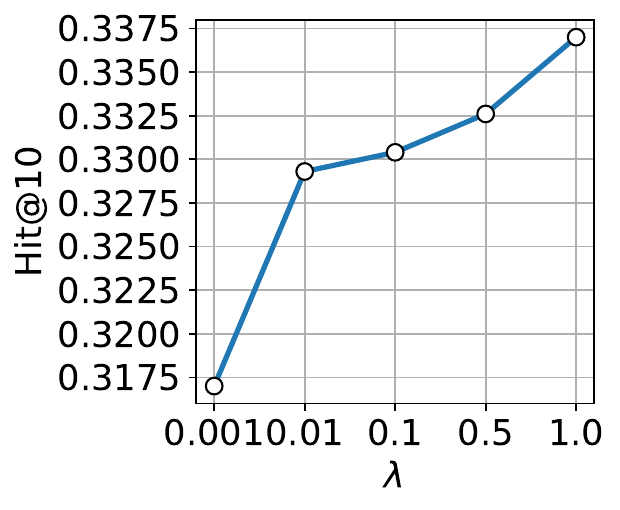}
        \caption{$\lambda$ on Christianity.}
    \end{subfigure}

    \caption{Hyperparameter sensitivity analysis: The influences of $\gamma$ and $\lambda$ on DRweibo, MGTAB and Christianity datasets.}
    \label{fig:param_analysis}
\end{figure}

In \Cref{fig:param_analysis}, the results show that the impact of $\gamma$ varies notably across tasks, reflecting their different levels of reliance on structural information. In DRWeibo (rumor detection), model performance peaks around $\gamma = 0.5$ and declines beyond that, suggesting that while graph learning helps, overemphasizing it may suppress the structure-agnostic patterns critical for classifying the root post. In MGTAB (social bot detection), the model benefits from relatively higher values of $\gamma$ (up to 0.7), which aligns with the nature of node classification within large social networks, where local graph structural context is more informative. In contrast, on Christianity (information diffusion prediction), increasing $\gamma$ leads to consistent performance degradation, indicating that graph learning plays a limited role in retrieval-based link prediction, where structure-agnostic patterns in cascade sequences are more dominant.

In addition, we observe that $\lambda$ on DRWeibo and MGTAB, where performance peaks around $\lambda=0.5$ and drops thereafter. This confirms that over-weighting the kinetic-guided loss interferes with task-specific objectives, and highlights the importance of balancing domain knowledge with data-driven learning. Moreover, in the Christianity dataset, performance improves steadily with increasing $\lambda$, suggesting that the kinetic-guided loss is more beneficial in link prediction, where learning structured dynamics is particularly challenging.

\section{Conclusion}
In this paper, we proposed a robust propagation-aware representation learning (RPRL) framework that incorporates domain knowledge through kinetic modeling to enhance data-driven representations. The proposed method adopts a general architecture capable of addressing graph classification, node classification, and link prediction. Extensive experiments demonstrate that RPRL achieves state-of-the-art performance across all three tasks. Moreover, the model exhibits strong cross-dataset generalization, significantly outperforming existing baselines in zero-shot and few-shot transfer settings. These results highlight the effectiveness and robustness of our approach, and point toward a promising direction for developing foundation models for social media graph analytics.

\section*{Acknowledgment}
The Australian Research Council partially supports this work under the streams of Future Fellowship (No. FT210100624), Discovery Early Career Researcher Award (No. DE230101033), Discovery Project (No. DP240101108 and No. DP240101814), and Linkage Project (Grant No. LP230200892 and LP240200546).

\bibliographystyle{IEEEtran}
\bibliography{sample-base.bib}

\end{document}